# Macroscopic Polarization Rotation Induced by a Single Spin

Christophe Arnold[1], Justin Demory[1], Vivien Loo[1], Aristide Lemaître[1], Isabelle Sagnes[1], Mikhaïl Glazov[2], Olivier Krebs[1], Paul Voisin[1], Pascale Senellart[1] and Loïc Lanco[1,3]

1. Laboratoire de Photonique et de Nanostructures, CNRS UPR 20, Route de Nozay, 91460 Marcoussis, France
2. Ioffe Physical-Technical Institute of the RAS, 194021, St-Petersburg, Russia
3. Université Paris Diderot – Paris 7, 75205 Paris CEDEX 13, France

**Solid-state spins hold many promises for quantum information processing[1,2,3,4,5]. Entangling the polarization of a single photon to the state of a single spin would open new paradigms in quantum optics like delayed-photons entanglement[6], deterministic logic gates[7] or fault-tolerant quantum computing[8]. These perspectives rely on the possibility that a single spin induces a macroscopic rotation of a photon polarization[9]. Such polarization rotations induced by single spins were recently observed, yet limited to a few $10^{-3}$ degrees[10,11,12] due to poor spin-photon coupling. Here we report the amplification by three orders of magnitude of the spin-photon interaction, using a cavity quantum electrodynamics device. A single hole spin trapped in a semiconductor quantum dot is deterministically coupled[13] to a micropillar cavity. The cavity-enhanced coupling[14] between the incoming photons and the solid-state spin results in a polarization rotation by ±6° when the spin is optically initialized in either the up or down state. These results open the way towards a spin-based quantum network.**

Spin-photon entanglement has recently been demonstrated, between a photon emitted by a quantum emitter and the spin degree of freedom of the same emitter[3,4,5]. A more scalable venue to spin-photon interfacing is to make use of the rotation of optical polarization (so-called Faraday or Kerr rotation) induced by a single spin placed at the center of a cavity-quantum electrodynamics (QED) device. This approach allows interfacing the spin with a photon generated by an external source, opening new possibilities in quantum optics[6,7,8,9,15,16,17] such as the engineering of temporally-delayed photon-photon interactions[6]. While Faraday or Kerr polarization rotation in a magnetised medium is routinely used for magnetic material characterisation[18], observations of Kerr rotation induced by a single spin were reported only recently[10,11,12], with rotation angles in the few $10^{-3}$ degrees range. In this work, a resident hole spin in a semiconductor quantum dot-pillar cavity device is initialized[19] and measured using resonant pump and probe beams: a Kerr rotation of several degrees is obtained. Still larger polarization rotations are found to be achievable with realistic cavity-QED devices, allowing the implementation of quantum functionalities with single photons interfaced to stationary spin qubits.

We study a single hole spin in a quantum dot (QD) efficiently coupled to the mode of a micropillar cavity (Fig. 1a). The pillar device, deterministically fabricated with the in-situ lithography technique[13], presents an optimal QD-cavity coupling strength, mostly polarization-degenerate optical modes, and an efficient coupling with external beams[14,20]. The quantities describing this device, sketched in Fig. 1b, are the QD-cavity coupling strength $g$, the QD dephasing rate $\gamma$, and the total damping rate $\kappa = \kappa_1 + \kappa_2 + \kappa_s$ (with $\kappa_1$, $\kappa_2$ and $\kappa_s$ the top mirror, bottom mirror and side leakage rates). These parameters are combined into two independent figures of merit, the top mirror output coupling efficiency $\kappa_1/\kappa$ and the device cooperativity $C = g^2/\kappa\gamma$.

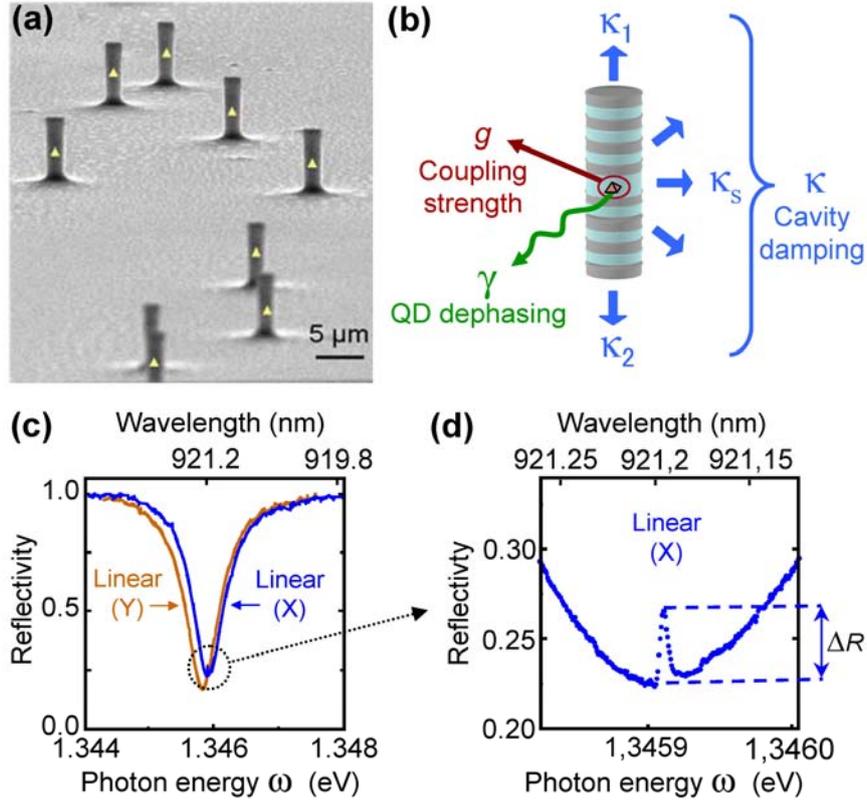

**Figure 1 | Quantum dot – pillar cavity device.**
**a.** Scanning electron microscope view of various micropillar devices. Yellow triangles denote the positions of the deterministically-coupled quantum dots. **b.** Sketch of the QD-micropillar device, with the corresponding cavity-QED parameters. **c.** Device reflectivity $R$ as a function of the laser photon energy, for two orthogonal linear polarizations $X$ and $Y$. The two modes are nearly polarization-degenerate. **d.** Zoom on the QD-induced transition, which is resonant at T=20 K with the $X$-polarized mode.

Preliminary device characterization is performed using coherent reflection spectroscopy (see Methods)[14,20,21]. The device reflectivity $R$ is displayed in Fig. 1c for two orthogonal polarizations (noted $X$ and $Y$) corresponding to the minor-axis and major-axis of the slightly elliptic micropillar. For both polarizations, the reflectivity spectrum displays a Lorentzian cavity dip with $\kappa$ = 630 μeV full-width at half-maximum, corresponding to a quality factor Q=2140. The mode spitting (90 μeV) is much smaller than $\kappa$, so the modes are close to polarization-degeneracy. The top- and bottom-mirror output-coupling efficiencies are estimated to be $\kappa_1/\kappa = \kappa_2/\kappa \approx 0.4$ corresponding to a low sidewall leakage contribution $\kappa_s/\kappa \approx 0.2$ [22]. This constitutes a substantial improvement compared to the device recently used to demonstrate a nonlinear optical response to few-photon pulses[14], where $(\kappa_1+\kappa_2)/\kappa$ was around 0.16. Finally, Fig. 1d displays a zoom centered on the $X$-polarized cavity dip, showing a narrow peak evidencing the efficient interaction with the QD transition: the resonantly excited dipole generates an optical field that interferes coherently with the exciting field. As will be shown, the observed transition corresponds to a QD which is charged with a resident hole thanks to the sample residual $p$ doping.

Figure 2a presents several spectra, centered on the QD transition, measured with linearly ($X$) polarized excitation and for different incident powers $P_0$. The QD resonance progressively disappears when the intracavity photon number increases in the cavity: this is the optical nonlinearity effect resulting from the saturation of the QD transition, as recently shown in Ref. 14. Fig. 2b displays similar spectra measured under left-handed (*L*) circularly-polarized excitation: a similar trend is observed, i.e. a gradual disappearance of the QD-induced peak with increased power, yet for much lower powers. The QD-induced reflectivity Δ*R*, defined as the absolute reflectivity subtracted by its value away from the QD resonance peak (see Fig. 1d), is plotted in Fig 2c as a function of the incident power $P_0$. It shows that the threshold for the peak disappearance is two orders of magnitude lower for the circular polarization. As discussed below, this lower threshold is a signature that optical initialization of a resident-hole spin is occurring, as reported in Ref. 19.

The hole ground state presents two spin configurations denoted $|\Uparrow\rangle$ and $|\Downarrow\rangle$. The positive trion (two holes and an electron) formed after photon absorption is in either $|\Uparrow\Downarrow\uparrow\rangle$ or $|\Downarrow\Uparrow\downarrow\rangle$ state ($|\downarrow\rangle$ denotes the electron spin state). Whenever the hole is in state $|\Uparrow\rangle$, thanks to the polarization selection rules illustrated in Fig 2d, optical pumping with *L*-polarized photons induces cycling transitions between the $|\Uparrow\rangle$ and $|\Uparrow\Downarrow\uparrow\rangle$ states. Contact hyperfine interaction with nuclei may occasionally induce an electron spin-flip leading to state $|\Downarrow\Uparrow\downarrow\rangle$, and eventually to state $|\Downarrow\rangle$ after spontaneous emission. Because the hole spin-flip time $T_1^{(hole)}$ is much larger than the trion spin-flip time $T_1^{(trion)}$, this results in a stationary regime where the hole is initialized in state $|\Downarrow\rangle$[19]. The QD-induced peak then disappears from the reflectivity spectra, due to the $|\Downarrow\rangle$-$|\Downarrow\Uparrow\downarrow\rangle$ transition being transparent to incident *L*–polarized light. In summary, in circular polarization QD transparency is induced at low excitation power by the spin pumping effect, whereas in linear polarization it is induced by the saturation of both trion transitions, arising at higher excitation power.

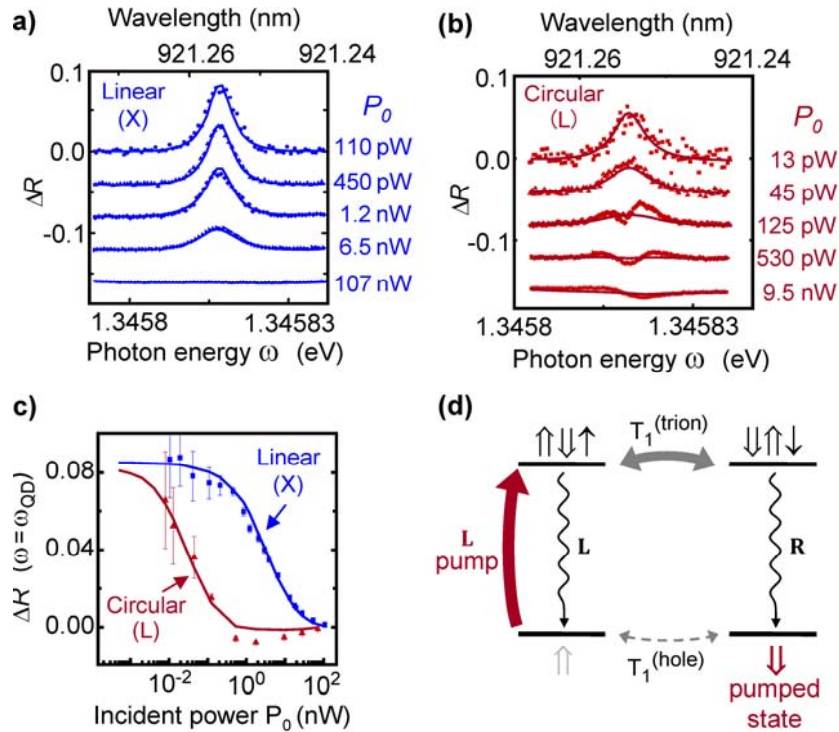

**Figure 2 | Optical pumping of a single spin** (in all panels symbols are used for experimental measurements and solid lines for numerical simulations). **a.** (resp. **b**.) Reflectivity spectra centered on the QD-induced peak, for various values of the incident power $P_0$, using linear (resp. circular) polarization. **c.** Reflectivity variation at the QD transition energy ω=$ω_{QD}$, for circularly and linearly polarized light, as a function of the incident power. **d.** Principle of hole spin pumping using a circularly-polarized pump.

As displayed in Figs 2a-c, numerical simulations have been performed by solving the system master equation [22]. Our observations are well reproduced with a high coupling strength $g = 15$ μeV and a low dephasing rate $γ = 2$ μeV. A ratio $T_1^{(hole)}/T_1^{(trion)} = 200$ is obtained from the fit of the power difference between the disappearance thresholds under circular and linear polarization. These results show that, despite a finite polarization splitting of the cavity mode, optical initialization in a given spin state is achieved.

In the following, we measure the Kerr rotation induced by the spin pumped either in the $|⇑>$ or $|⇓>$ state. To do so, a weak probe beam with linear polarization state $(|L\rangle+|R\rangle)/\sqrt{2}$ is sent on the device. The reflected beam is then in the normalized state $(r_L|L\rangle+r_R|R\rangle)/\sqrt{|r_L|^2+|r_R|^2}$, with $r_L$ and $r_R$ the complex reflection coefficients for the $L$ and $R$ polarizations (see Fig. 3a). When the hole is in state $|⇓>$, the $L$ component of the probe beam is insensitive to the QD transition, whereas the $R$ component experiences a reflectivity shift induced by the QD transition: $r_R$ thus depends on the detuning between the QD transition energy $ω_{QD}$ and the probe photon energy $ω_{probe}$, while $r_L$ does not[7,9]. A symmetrical behaviour is expected when the hole is in state $|⇑>$. This results in a Kerr rotation of the probe beam polarization depending on the spin state. We note the corresponding output polarization states $|Ψ⇑>$ and $|Ψ⇓>$, for the spin states $|⇑>$ and $|⇓>$ [22].

The Kerr-rotation experimental setup is sketched in Fig. 3b. Two co-linear continuous-wave pump and probe beams, with photon energies $ω_{pump}$ and $ω_{probe}$, are focused on the upper surface of the micropillar. The pump beam polarization is chosen to be $L$ or $R$ to initialize the spin in either the $|⇑>$ or $|⇓>$ state. The probe beam is linearly ($X$) polarized and modulated at 100 kHz. The reflected beams are sent onto a polarizing beam-splitter (PBS) that separates the total reflected power into its horizontal and vertical components $P_H$ and $P_V$. These two components are measured with two avalanche photodiodes, and a lock-in amplifier filters the signal contribution arising from the probe. The Kerr rotation angle is then deduced from the measured photodiode contrast, $(P_V-P_H)/(P_V+P_H)$. In the absence of polarization rotation, a zero contrast is observed. On the contrary, a positive (resp. negative) contrast will be observed in the case of a clockwise (resp. counter-clockwise) polarization rotation.

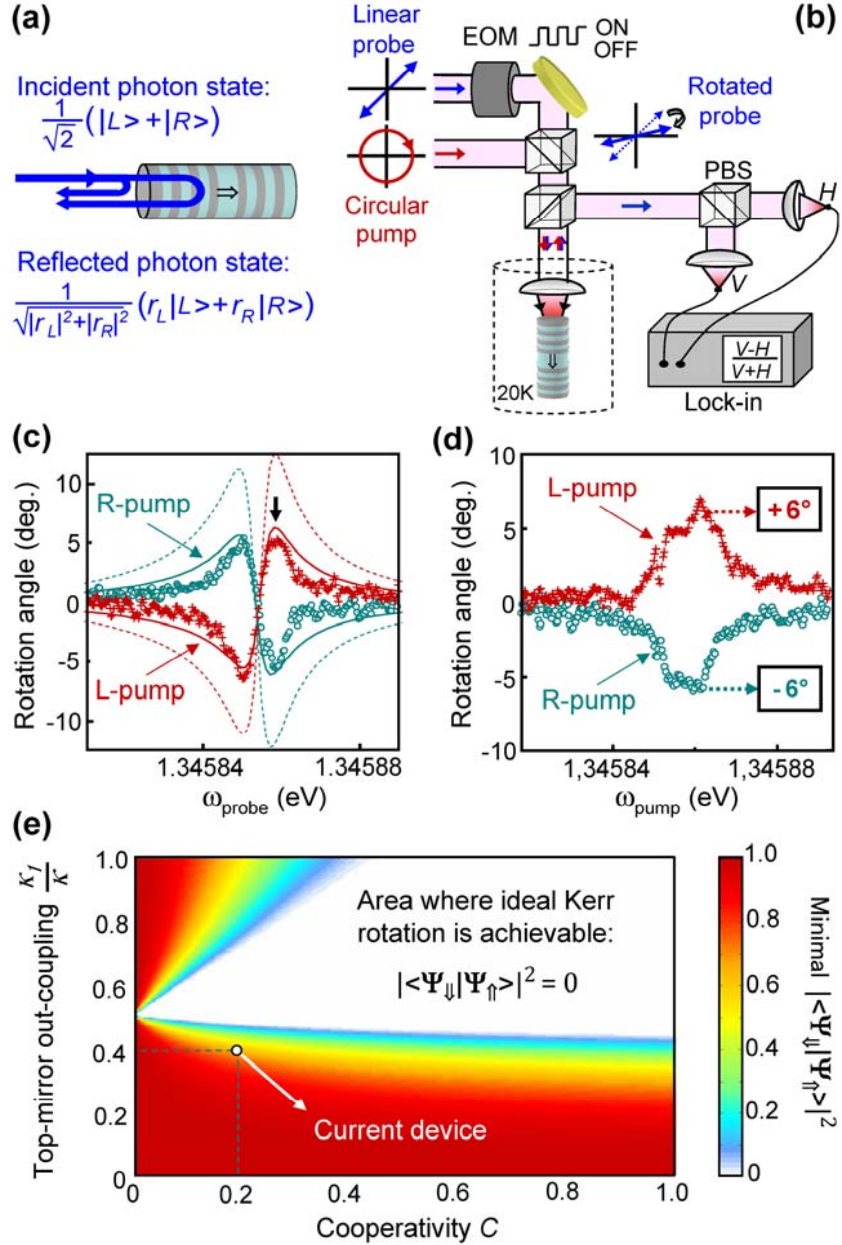

**Figure 3 | Macroscopic Kerr rotation induced by a single spin in a cavity-QED device.**

**a.** Polarization states for the incident and reflected beam. The reflected beam results from the interference of two contributions: direct reflection and light injected into and re-extracted from the cavity. **b.** Simplified scheme of the experimental setup used for spin optical pumping and Kerr rotation measurements on a QD-pillar cavity device. **c.** Kerr rotation angle as a function of $\omega_{probe}$, with $\omega_{pump}$ fixed at 1.345857 eV (symbols: experimental data, solid line: theoretical fit with partial spin initialization, dashed line: theoretical prediction with perfect spin initialization). **d.** Kerr rotation angle as a function of $\omega_{pump}$, with $\omega_{probe}$ fixed at a maximum of Kerr rotation (thick vertical arrow in panel c). **e.** Minimal value of $|\langle\Psi_\Downarrow|\Psi_\Uparrow\rangle|^2$ achievable for a given set of device parameters $C$ and $\kappa_1/\kappa$; the ideal configuration $\langle\Psi_\Downarrow|\Psi_\Uparrow\rangle=0$ can be obtained for a large range of parameters.

Fig. 3c presents the Kerr rotation angle measured for both *L* and *R* pump polarizations, for a pump beam in resonance with the QD transition. A dispersive shape centered on $\omega_{QD}$ is observed for the Kerr rotation signal as a function of $\omega_{probe}$, as previously reported in Ref 10. However, here the Kerr rotation is macroscopic, with a maximum rotation angle of ± 6°. This improvement by three orders of magnitude can quantitatively be accounted for by cavity quantum electrodynamics effects in an optimal cavity geometry. Two main features account for this huge amplification. The first one is the increase of the spin-photon interaction induced by the optical confinement. This enhancement is governed by the device cooperativity $C = g^2/\kappa\gamma \approx 0.2$, which describes how efficiently a single two-level system modifies the optical properties of the confined mode. The second feature, specific to pillar-cavity devices, is the efficient interference between the directly-reflected light and the light injected into and re-extracted from the cavity (see Fig. 3a). This interference is governed by the top-mirror output-coupling efficiency $\kappa_1/\kappa \approx 0.4$.

The expected Kerr rotation angle is calculated using the QD and cavity parameters obtained from fitting the experimental data for spin pumping. The calculated rotation angle for perfect spin initialization fidelity is presented in dashed lines in Fig. 3c. The expected maximum rotation angle is 12°, twice larger than the experimental one, indicating an imperfect spin pumping in the pump-probe experiment. Indeed, to obtain a good signal to noise ratio, a probe power $P_{probe}$=110 pW is used, representing 10% of the pump power $P_{pump}$=1.3 nW. This leads to a non negligible de-pumping of the spin state. A good agreement with the experimental data is obtained with a partial spin initialization, where the spin is in the desired state with a probability 0.75. As expected, the Kerr rotation is fully reversed when the pump polarization is set to $R$, corresponding to a spin pumping in the $|\Uparrow>$ state. Further evidence that this rotation arises from a single spin is obtained by measuring the Kerr rotation with varying $\omega_{pump}$, at a fixed $\omega_{probe}$. Fig. 3d shows that the maximal Kerr rotation angle is obtained for $\omega_{pump}=\omega_{QD}$, i.e. when the optical spin pumping is the most efficient.

Most theoretical works concerning the use of a spin in a cavity, to implement new functionalities in quantum optics and quantum computing, rely on the assumption of large Kerr rotations. Our results show for the very first time that such macroscopic rotation induced by a single spin can experimentally be obtained. We now discuss how our results could be further improved by optimizing the device. An important figure of merit for quantum applications is the scalar product $<\Psi_\Downarrow|\Psi_\Uparrow>$ between the two possible output polarization states. The ideal case, where $<\Psi_\Downarrow|\Psi_\Uparrow>=0$, would indeed allow a perfect mapping between the spin state and the photon polarization state [23]. Indeed, if $|\Psi_\Uparrow>$ and $|\Psi_\Downarrow>$ are orthogonal they can be converted into $|H>$ and $|V>$ states using a set of half-wave and quarter-wave plates. In such a case a strong projective measurement can be obtained with a single reflected photon, whose detection in the $H$ (respectively, $V$) detector will automatically project the spin in state $|\Uparrow>$ (respectively, $|\Downarrow>$).

For each device, there is a minimal value of $|<\Psi_\Downarrow|\Psi_\Uparrow>|^2$ which can be achieved through a proper tuning of the QD-cavity detuning and probe photon energy. Fig. 3e shows the calculated minimal value of $|<\Psi_\Downarrow|\Psi_\Uparrow>|^2$ achievable, as a function of the device cooperativity $C$ and top-mirror output-coupling $\kappa_1/\kappa$; the current device with $\kappa_1/\kappa \approx 0.4$ and $C = 0.2$ is indicated by the white circle. The ideal situation $<\Psi_\Downarrow|\Psi_\Uparrow>=0$ can be reached for a large range of realistic values of $C$ and $\kappa_1/\kappa$, for example with the same cooperativity $C = 0.2$ but with $\kappa_1/\kappa \approx$ 0.6. This can be obtained by increasing the number of pairs in the bottom Bragg mirror, thus decreasing $\kappa_2$

$/\kappa$, and correspondingly increasing $\kappa_1/\kappa$. A perfect quantum non-demolition measurement of the spin state, using a single reflected photon, would thus be possible with such a device. Furthermore, provided that the spin can be initialized in a well-defined coherent superposition of $|\Uparrow\rangle$ and $|\Downarrow\rangle$, using for instance optical[24] or microwave pulses[25], a maximally-entangled state $(|\Downarrow,\Psi_\Downarrow\rangle+|\Uparrow,\Psi_\Uparrow\rangle)/\sqrt{2}$ could be produced between the spin and the reflected photon. This is a key building block for a large number of theoretical proposals at the heart of a future solid-state quantum network.

## METHODS SUMMARY

**Device fabrication.** A GaAs/Al$_{0.9}$Ga$_{0.1}$As microcavity embedding self-assembled InAs quantum dots is grown by molecular beam epitaxy on a GaAs substrate. The bottom and top Bragg mirrors have 24 and 20 pairs, respectively, and present equal reflectivities (ie $\kappa_1 = \kappa_2$). The in-situ lithography technique is implemented to define pillars centered on single QDs. A red laser beam is first used to excite and monitor the QD emission, so as to measure its spatial position with 50 nm accuracy. A green laser beam is then used to expose, at the QD position, a layer of photoresist which is subsequently used as a mask for inductively coupled plasma etching of the micropillar. The device selected in this work is charged with a resident hole and has a diameter of 2.1 μm.

**Experiments.** The sample is maintained at T=20K inside a helium-vapor cryostat, together with an aspheric lens and cryogenic nanopositioners. The reflectivity is deduced from the reflected and incident powers measured with avalanche photodiodes, and then normalized to unity when the laser photon energy is far from the cavity mode resonance. No magnetic field has been applied. Spin initialization and Kerr rotation measurements were performed using liquid-crystal variable waveplates to compensate the polarization distortions induced by the various components along the optical paths.

**Acknowledgments.** The authors acknowledge A. Dousse and J. Suffczynski for the pillar sample and E.L. Ivchenko for valuable discussions. This work was partially supported by the French ANR MIND, ANR QDOM, the ERC starting grant 277885 QD-CQED, the CHISTERA project SSQN, the French Labex NANOSACLAY, and the RENATECH network.